\documentclass[aip,
 amsmath,amssymb,
reprint,%
]{revtex4-1}

\usepackage{graphicx}
\usepackage{dcolumn}
\usepackage{bm}

\usepackage[utf8]{inputenc}
\usepackage[T1]{fontenc}
\usepackage{mathptmx}
\usepackage{etoolbox}

\makeatletter
\def\@email#1#2{%
 \endgroup
 \patchcmd{\titleblock@produce}
  {\frontmatter@RRAPformat}
  {\frontmatter@RRAPformat{\produce@RRAP{*#1\href{mailto:#2}{#2}}}\frontmatter@RRAPformat}
  {}{}
}%
\makeatother
\begin{document}

\preprint{AIP/123-QED}

\title[]{Spectral characterization of a SPDC source with a fast broadband spectrometer}
\author{Brianna Farella}
 \affiliation{ 
Brookhaven National Laboratory, Upton NY 11973, USA\
}
\author{Gregory Medwig}
\affiliation{ 
Brookhaven National Laboratory, Upton NY 11973, USA\
}%
\author{Raphael A. Abrahao}
\email{rakelabra@bnl.gov}
\affiliation{ 
Brookhaven National Laboratory, Upton NY 11973, USA\
}%
\author{Andrei Nomerotski}
\email{anomerotski@bnl.gov}
\affiliation{ 
Brookhaven National Laboratory, Upton NY 11973, USA\
}%

\date{\today}

\begin{abstract}
Knowing the properties of the single photons produced in a Spontaneous Parametric Down-Conversion (SPDC) source can be crucial for specific applications and uses. In particular, the spectral properties are of key relevance. Here, we investigate a commercial SPDC source using our fast broadband spectrometer. Our analysis is a valid method for other SPDC sources, as well as other single-photon generation techniques, thus providing a good example of how to use this spectrometer design. We calibrate the spectrometer using known lines of the argon emission spectrum. We show that the two down-converted photons from the SPDC source have different spectral properties depending on the pump power, and in which condition we measured spectrally similar down-converted photons. Lastly, we were able to reconstruct and investigate the spectral information for the pump photon.
\end{abstract}

\maketitle

\section{Introduction}
\label{sec:intro}

With emerging optical quantum technologies, so arise commercial single-photon sources. Albeit practical, these sources need to match the demands of specific application uses. Spontaneous Parametric Down-Conversion (SPDC) is one of the most common methods to generate single photons. Here, we report an important spectral analysis of the generated single photons from a Thorlabs SPDC source, which is based on a ppKTP crystal. The methods presented here are also valid for other single-photon sources and demonstrate the versatility of our fast spectrometer design. Although these commercial SPDC sources are convenient, as an off-the-shelf solution for some users, they provide limited versatility and restrict the ability to control the properties of the desired photon emission. Typically, the power of the pump laser is the only available degree of freedom.

Spontaneous Parametric Down-Conversion (SPDC) is a probabilistic nonlinear-optical~\cite{book_boyd} process in which a pump photon is absorbed with a subsequent emission of a photon pair~\cite{SPDC_general,PRA_Kwiat99,PRL_Kwiat95,PRL_Oxford_spdc2008,spdc_agata_PRA2016,spdc_branczyk2011engineered,spdc_grice_PRA2001,spdc_grice_PRA97}. In order for this effect to take place, two conditions need to be satisfied: conservation of momentum, which in turn leads to the phase-matching condition, and conservation of energy, i.e. $\hbar\omega_{pump}=\hbar\omega_{signal}+\hbar\omega_{idler}$. The two photons produced are often called signal and idler, and the wavelengths of signal and idler photons are anti-correlated.

Knowing the spectral properties of single photons is essential for many practical applications, including determining the possibility of the Hong-Ou-Mandel effect \cite{HOM_effect,Bouchard_HOM_review,Sensors2020_Nomerotski,PRA_Jordan2022}, optical quantum information processing \cite{kok2007linear,kok2010introduction,cnot2003UQ,KLM2001,Geoff_P_review}, quantum astronomy \cite{Stankus2022,arxiv_BNL_qtelescopes,Zhi2022,Keach2022, Nomerotski20,fast_spectrometer} and many other optical quantum technologies. This analysis is of high value for our future experiments and the broad quantum optics community. The spectrometer presented here can also be used to characterize other kinds of single-photon sources, including quantum dots~\cite{qd_review_senellart2017,qd_gazzano2013,qd_somaschi2016,qd_santori2002,qd_he2013,qd_patel2010,qd_lenzini2017active} and four-wave mixing~\cite{fwm_mcmillan2013,fwm_paesani2020,fwm_PRA2004,fwm_PRL2005}.

\section{SPDC photon source}
For our experiment, we used a Thorlabs SPDC810 Spontaneous Photon Pair Source (Correlated Photon-Pair Source). The pump laser wavelength is 405 nm, and the two down-converted photons each have a wavelength of about 810 nm. One of these down-converted photons is labeled signal and the other is labeled idler, both of which have their bulkhead fiber port to interface with. 
We used single-mode fibers, each approximately 1 meter in length, to connect the signal and idler photons to the two channels on the spectrometer.

The pump laser power can be set to a specific value, with a maximum setting of 150 mW. Using superconducting nanowire single-photon detectors (SNSPD) with high detection quantum efficiency, we were able to obtain single-photon rates (one arm of the SPDC) exceeding 450k detections per second at maximum power.

\section{Spectrometer}

\begin{figure}[]
\centering
\includegraphics[width=.4\textwidth]{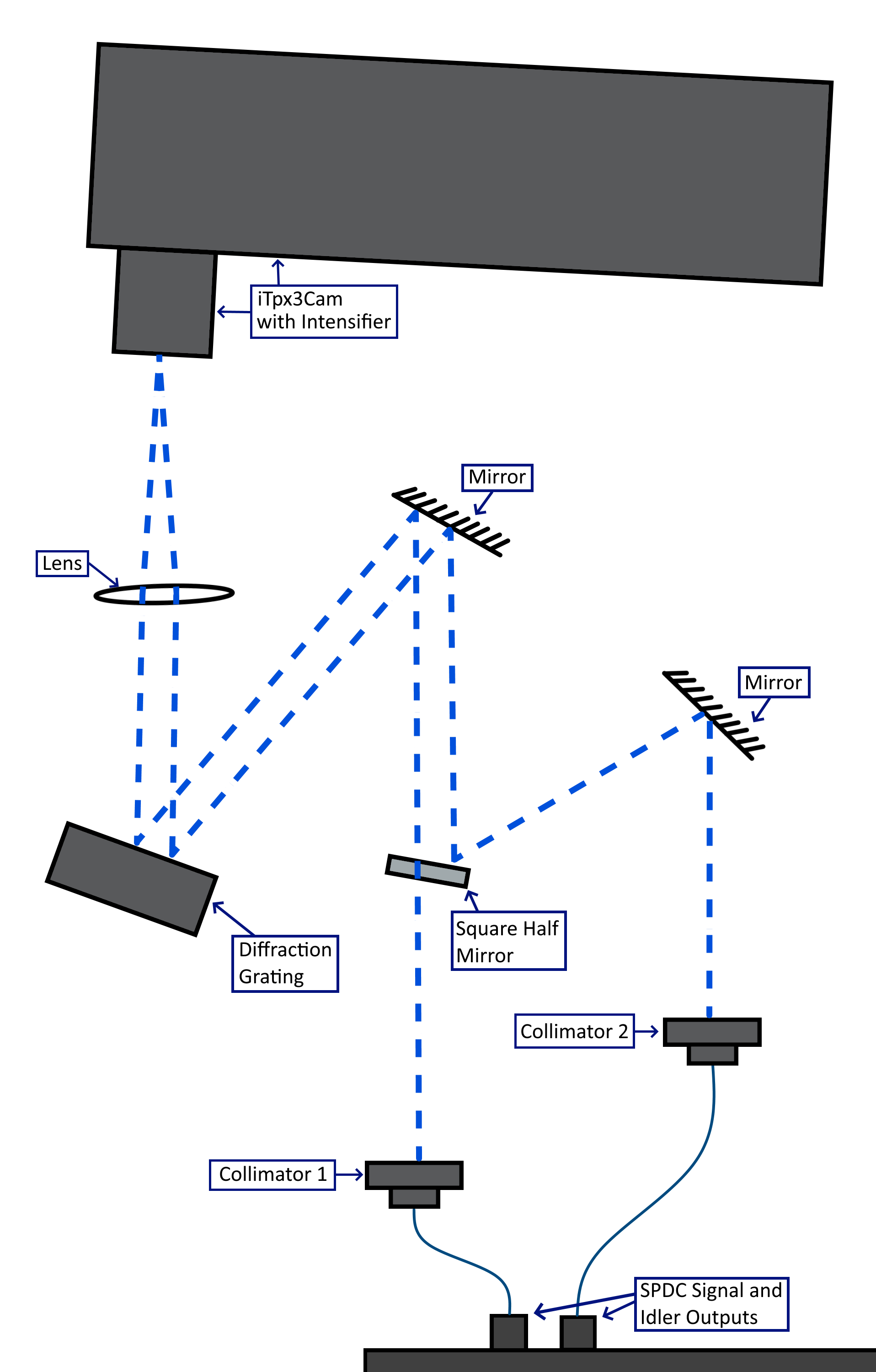}
\caption{Diagram showing the spectrometer setup. The signal and idler photons of the SPDC source are connected to the spectrometer using single-mode optical fibers. The dashed blue lines show the paths of the two beams. A series of mirrors and a half mirror are employed to send the beams to a diffraction grating (1200 lines/mm), and then focused ($f$=100 mm) and entering the intensified Timepix3 Camera. \label{fig:Spectrometer}}
\end{figure}

Figure \ref{fig:Spectrometer} shows a schematic diagram of our experimental setup. The signal and idler of the SPDC source are coupled to respective single-mode optical fibers. Then, these photons are guided to our spectrometer with their optical paths indicated with dashed lines in the figure. The photons are collimated into two parallel beams of about 2 mm diameter and are reflected off mirrors to send the beams to a diffraction grating (1200 lines/mm) and then focused ($f$=100 mm) onto the input window of the intensified Timepix3 camera, Tpx3Cam \cite{Nomerotski2019}, a similar spectrometer design as references~\cite{Yingwen2020,Zhang2021}.

As the system is single-photon sensitive, it is very susceptible to sources of light, even if they are weak. To prevent unwanted light from entering the spectrometer, the setup is placed in a dark enclosure in a dark room.

\subsection{Intensified fast camera Tpx3Cam}
Imaging of single photons for these measurements was done with a data-driven time-stamping camera, called Tpx3Cam \cite{timepixcam, tpx3cam, Nomerotski2019, Nomerotski2023, ASI}. This camera employs a silicon sensor with back-side illumination, which provides a high quantum efficiency (QE) in the range of 400-950~nm \cite{Nomerotski2017}. The sensor is attached to the Timepix3 \cite{timepix3} application-specific integrated circuit (ASIC) via bump-bonding. The chip comprises $256 \times 256$ pixels of $55 \times 55 $ $\mu$m$^2$. The time-of-arrival (ToA) is measured using the in-pixel electronics, which processes the incoming signals for hits that cross a predefined threshold of around 600 electrons, with a precision of 1.56 ns. Another signal property, measured simultaneously with ToA for each detection, is the time-over-threshold (ToT), which is related to the energy deposited in each pixel. The ToT values are encoded with the ToA values as time codes. The readout of Timepix3 is data-driven, with a pixel deadtime of microseconds, which allows for multi-hit functionality of every pixel independently from others and with fast bandwidth of 80 Mpix/sec \cite{heijdenSPIDR}.

For the single-photon-sensitive operation, the signal is amplified with the addition of Cricket$^{\rm{TM}}$~~\cite{Photonis}, a integrated device for the image intensifier, which in turns is a vacuum apparatus composed by a photocathode, a micro-channel plate, and a P47 scintillator. The scintillator has a rise time of about 7~ns and maximum emission at 430 nm \cite{Winter2014}. The image intensifier operates at a high gain of about $10^6$, which ensures that a detected single photon produces a signal well above the camera threshold.  The quantum efficiency of the intensified camera is determined primarily by the intensifier photocathode with a variety of photocathodes available, covering the spectral range 120~--~900~nm \cite{Nomerotski2019, Photonis}. The intensifier used for our measurements had a hi-QE-red photocathode with QE of about 20\% in the range of 550 - 850~nm \cite{Orlov2016} and a chevron MCP with an improved detection efficiency close to 100\% \cite{Orlov2018}.

\subsection{Camera data post-processing}

The photons appear in the camera as small collections of hit pixels. In the post-processing phase, these pixels are organized into clusters within a predefined time window employing a recursive algorithm \cite{tpx3cam}. The ToT information, measured independently for each pixel, can be used for centroiding to determine the coordinates of single photons, improving spatial resolution \cite{Hirvonen2017}. The timing of the photon is typically estimated by using the ToA of the pixel with the largest ToT in the cluster, but more complex algorithms can be used as well.
The time-of-arrival is then adjusted for the time-walk, caused by the different front-end electronics temporal response, which depends on the amplitude of the input signal \cite{Ianzano2020}.

\section{Experimental Procedure}

We describe below the experimental procedure used for characterization of the SPDC source including the spectrometer calibration using the argon spectral lines.

\subsection{Spectrometer Calibration}

To use the spectrometer for precision spectral measurements, it needs first to be calibrated. In our case, we needed to verify the precise relative calibration of two spectrometers.  To calibrate them, we used an argon thermal lamp. The light from the Ar lamp is collected into a single-mode optical fiber before sending it to a fiber-coupled beamsplitter, so that both ends can be connected to the two channels of the spectrometer setup. 

\begin{figure}[]
\centering
\includegraphics[width=0.47\textwidth]{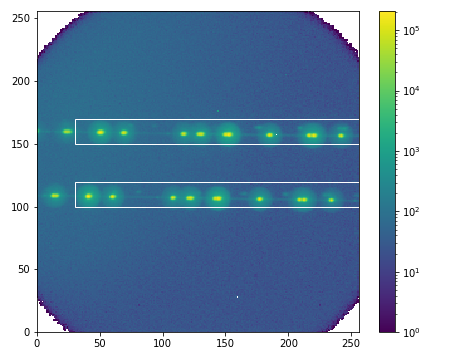}
\caption{Two argon spectra recorded with the Tpx3Cam, corresponding to the different emission lines of argon. Each region of interest can be interpreted as a result of an independent spectrometer, called top and bottom.}
\label{fig:SPDC_Ar_Spec}
\end{figure}

The dual argon spectra recorded in the fast camera are shown in Fig.~\ref{fig:SPDC_Ar_Spec}. As expected, we observe a discrete spectrum corresponding to the different emission lines of argon. The total spectrum range was about 120~nm, from 740~nm to 860~nm. This range was selected with the alignment of the setup. The typical duration of the calibration dataset was 10~s. 

\begin{figure}[]
\centering
\includegraphics[width=0.5\textwidth]{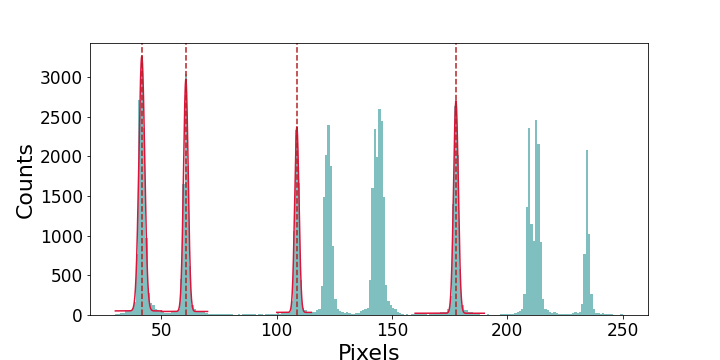}
\caption{X-projection of the argon spectrum selected in the bottom box in Fig.~\ref{fig:SPDC_Ar_Spec}. Argon spectrum x-projection as a 1D histogram. Singlet spectrum lines 763.51 nm, 772.38 nm, 794.82 nm, and 826.45 nm are fit with respective Gaussian curves (in red) with their center point marked by dashed red lines.}
\label{fig:SPDC_Ar_Spec_hist}
\end{figure}

Two regions of interest in \textit{x} and \textit{y} were selected in the camera, which represent the spectrum observed from each optical path, indicated with boxes in Fig.~\ref{fig:SPDC_Ar_Spec}. Fig.~\ref{fig:SPDC_Ar_Spec_hist} shows a x-projection of the data corresponding to one of the boxes. 

\begin{figure}[!htbp]
\centering
\includegraphics[width=0.45\textwidth]{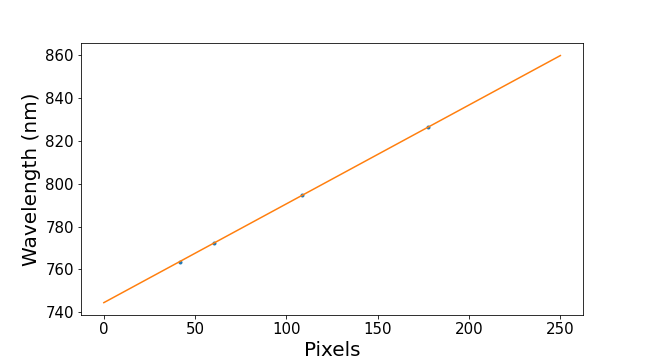}
\caption{Center points of the Gaussian curves plotted and fit with a line to produce a conversion equation for the bottom region of interest of Fig.~\ref{fig:SPDC_Ar_Spec}.}
\label{fig:Ar_spec_conv}
\end{figure}

Singlet spectrum lines 763.51 nm, 772.38 nm, 794.82 nm, and 826.45 nm are fit with respective Gaussian curves shown in red with their center point marked by dashed lines. The sigmas of the Gaussian fits are consistently about 0.15~nm indicating good spectral resolution of the spectrometers. To determine the spectrometer scale we plot the center points of the Gaussian curves and fit them with a line to produce a conversion scale as shown in Fig.~\ref{fig:Ar_spec_conv} for one of the spectrometers.

The measured conversion scales for the two spectrometers (top and bottom regions of interest of Fig.~\ref{fig:SPDC_Ar_Spec}) were equal to 0.462 ± 0.002 pix/nm (bottom) and 0.467 ± 0.003 pix/nm (top), which demonstrates a good uniformity of the setup. The linear fits are y = 0.462x + 744.416 (bottom), and y = 0.467x + 739.880 (top). The calibration is then applied to the SPDC datasets to determine the spectral distributions of the idler and signal photons.


\subsection{SPDC spectral analysis}

After calibration with the argon lamp, we used the SPDC source, using SMFs to send the light to the spectrometer. We connected one SPDC source port to one channel of the spectrometer and the other SPDC port to the other channel.

\begin{figure}[]
\centering
\includegraphics[width=0.44\textwidth]{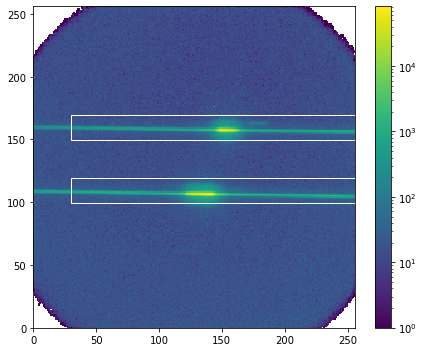}
\caption{Camera recording of the SPDC data for idler and signal photons in two spectrometer channels as recorded with Tpx3Cam. For this figure, the pump power was set to 100 mW.}
\label{fig:SPDC_spectrum}
\end{figure}

We then took three datasets, with each data taking duration of 10~s, setting the SPDC pump power to 50, 100, and 150 mW. An example of the camera recording of a SPDC spectrum is presented in Fig.~\ref{fig:SPDC_spectrum}.

\begin{figure}[!htbp]
\centering
\includegraphics[width=0.5\textwidth]{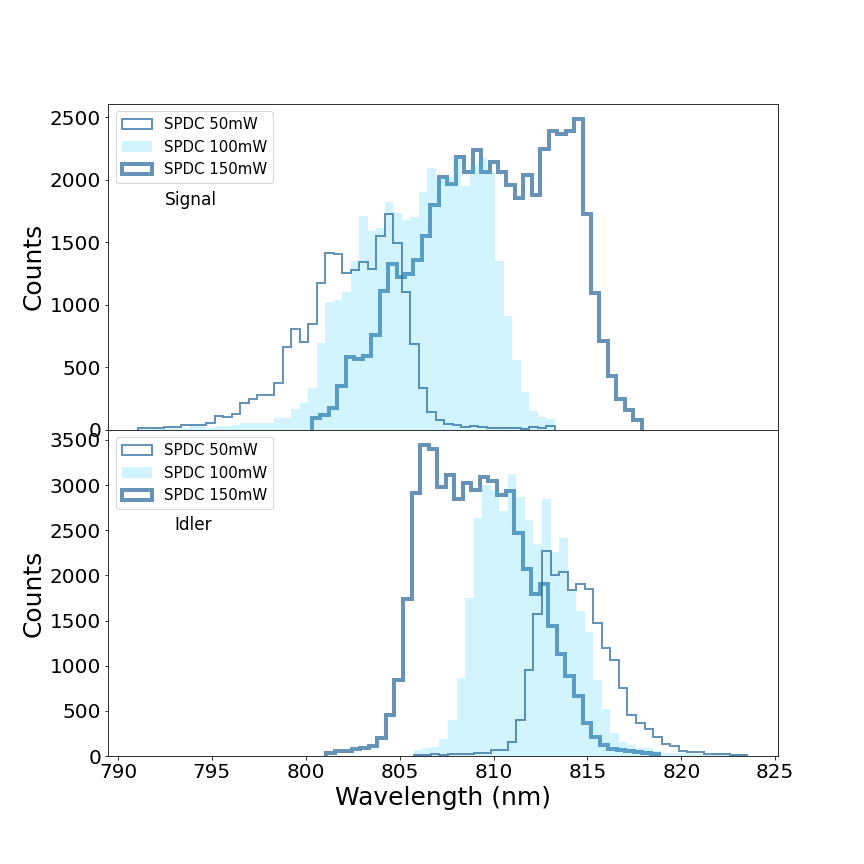}
\caption{Histogram showing wavelengths of the signal and idler for the three power settings. The data taking for each power level was set to 10 seconds.
\label{fig:sig_idl_pos}}
\end{figure}

A procedure of separating the two SPDC spectra, the idler and signal photons, using region-of-interest boxes, as we did for the argon spectrum, was employed to produce the 1D histograms of the SPDC spectra. The spectra for three power levels of the SPDC source are shown in Fig.~\ref{fig:sig_idl_pos}.

\begin{figure}[!htbp]
\centering
\includegraphics[width=0.5\textwidth]{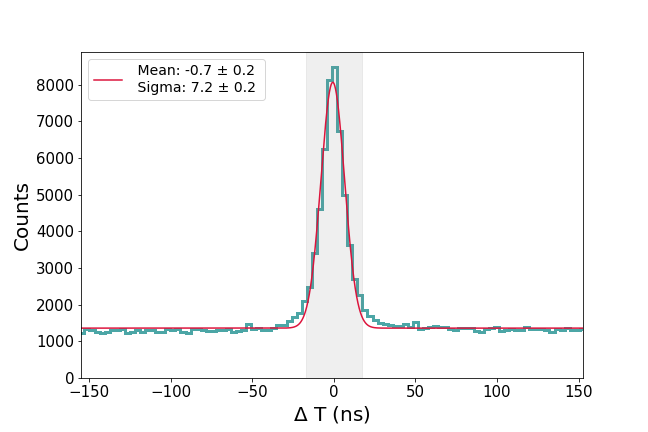}
\caption{Histogram showing the time difference between individual photons in the signal and idler channels, the grey band representing the selection used. The pump power setting was 150 mW.}
\label{fig:dt}
\end{figure}

Two photons from the SPDC process are produced simultaneously and, therefore, can be selected as coincident photon pairs from the two spectrometers. We devised an algorithm to look for such pairs, which was identifying the closest in time hits in the spectrometers \cite{Nomerotski20}.
Fig.~\ref{fig:dt} shows the time difference distribution between photons located in the signal and idler spectra for 150 mW pump. From this, we select differences which absolute values are less than 20~ns in order to encompass the entire peak corresponding to the coincidences. Fitting this with a temporal Gaussian profile, and assuming the sigma fitting as the temporal resolution, this spectrometer has $\approx$ 7 ns temporal resolution.

Table \ref{tab:FWHM} provides the median and FWHM values for the six spectral distributions. From Fig.~\ref{fig:sig_idl_pos}, our data shows that as the power increases, the separation between the signal and idler wavelengths decreases, with 50~mW having practically no overlap, 100~mW having partial overlap, and 150~mW having significant overlap. Another observation that can be made is that the center wavelength of the signal and idler curves appears to shift in opposite directions relative to each other as power increases, an expected effect due to the conservation of energy. We also noted that the idler spectral distribution is narrower than the signal distribution. 

\begin{table}[!htbp]
\begin{ruledtabular}
\begin{tabular}{cccc}
& power (mW) & median (nm)& FWHM (nm)\\
\hline
signal & 50 & 802.8 & 5.9\\
 & 100 & 806.5 & 8.3\\
 & 150 & 810.4 & 9.9\\
\hline
idler & 50 & 814.0 & 3.8\\
 & 100 & 811.4 & 5.1\\
 & 150 & 808.9 & 6.9\\
\end{tabular}
\end{ruledtabular}
\caption{\label{tab:FWHM} Median and FWHM values corresponding to the distributions seen in Figure \ref{fig:sig_idl_pos}.}
\end{table}

\begin{figure}[!htbp]
\centering
\includegraphics[width=0.4\textwidth]{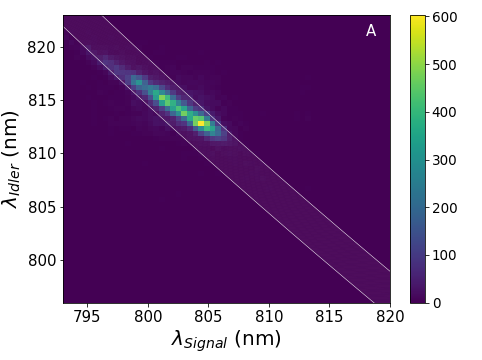}
\qquad
\includegraphics[width=0.4\textwidth]{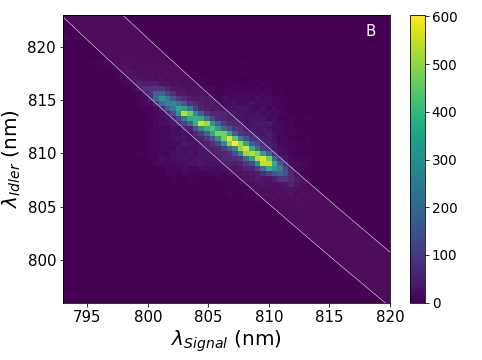}
\qquad
\includegraphics[width=0.4\textwidth]{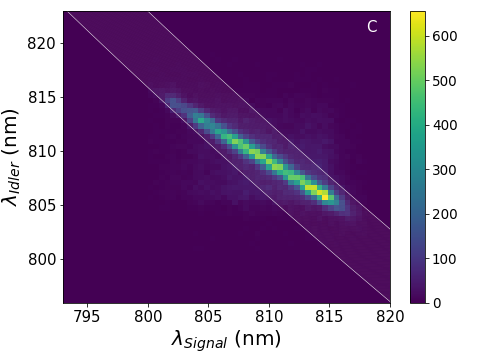}
\caption{Joint Spectral Intensity (JSI) for pump power levels of  50~mW (A), 100~mW (B), and 150~mW (C) as a function of the signal and idler photon wavelengths. The lines encompass a $\pm 2 \sigma$ band of the reconstructed pump photon wavelength as in Fig.~\ref{fig:pump_photon}.}
\label{fig:anticorrelation}
\end{figure}

In Fig.~\ref{fig:anticorrelation}, we plot the 2D correlations of the measured wavelengths for the matched photon pairs for three different pump power configurations. 
The graphs show a typical anti-correlation behaviour of wavelengths for signal and idler photons and could be interpreted as the Joint Spectral Intensity (JSI), defined by the conservation of energy and conservation of momentum, the latter corresponding to the phase matching condition in the down-conversion process. The conservation of energy can be mathematically written as Equation~\ref{equ:pump_wavelength}. 

\begin{equation} \label{equ:pump_wavelength}
\frac{hc}{\lambda _{pump}} = \frac{hc}{\lambda _{signal}} + \frac{hc}{\lambda _{idler}}
\end{equation}

The phase matching is normally related to material properties. In general, depending on material constraints, the JSI can be engineered to achieve a desired goal, such as some specific spectral characteristics for entangled down-converted photons \cite{SPDC_general}. 

We also note that our data could be affected by fluctuations in the pump wavelength and pump power, as well as thermal effects in the oven of the ppKTP crystal. These effects could make the allowed values for the down-converted photons to be spread over a larger interval in the JSI compared to the case of fixed pump photon energy and temperature stabilization.

\begin{figure}[!htbp]
\centering
\includegraphics[width=0.5\textwidth]{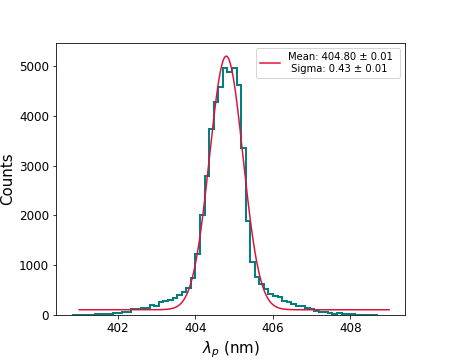}
\caption{Distribution of calculated pump photon wavelength with a Gaussian curve fit for the 150~mW pump power setting.} \label{fig:pump_photon}
\end{figure}

The wavelength of the pump photon can be calculated using the signal and idler wavelengths, as in Equation \ref{equ:pump_wavelength}. We spectrally reconstructed the pump photon and fit it with a Gaussian function, generally achieving a good agreement with the expected pump wavelength, 405~nm, see Fig.~\ref{fig:pump_photon} for the 150~mW pump power setting. 
We provide the central values and variation (rms) of the reconstructed pump photon wavelength for all three cases in Table \ref{tab:pump}.

\begin{table}[!htbp]
\begin{ruledtabular}
\begin{tabular}{ccc}
power (mW) & central value (nm)& rms (nm)\\
\hline
50 & 404.13 & 0.27\\
100 & 404.47 & 0.33\\
150 & 404.80 & 0.43\\
\end{tabular}
\end{ruledtabular}
\caption{\label{tab:pump} Central values and variation (rms) of the reconstructed pump photon wavelength for three pump power settings.}
\end{table}

\begin{figure}[!htbp]
\centering
\includegraphics[width=0.4\textwidth]{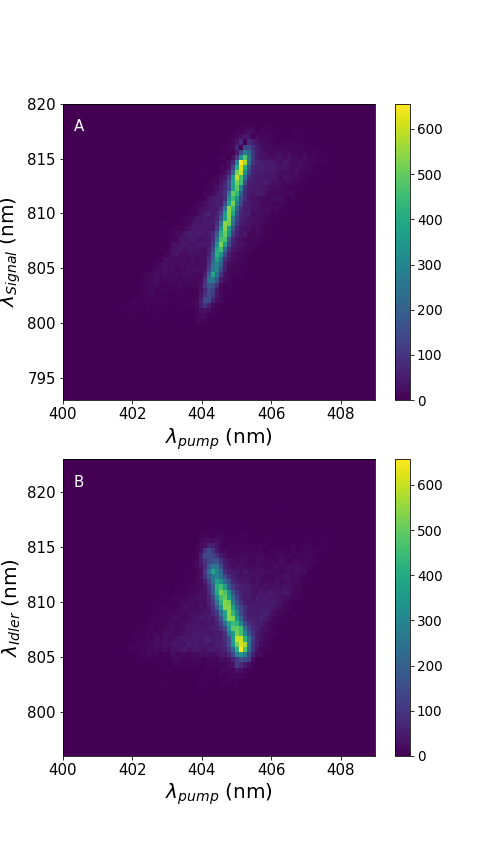}
\caption{Correlation between the individual down-converted photon wavelength and the calculated pump photon energy for two cases of signal (A) and idler (B) photons for the 150~mW pump power setting.}
\label{fig:channel_v_pump_photon}
\end{figure}

We further plot in Fig.~\ref{fig:channel_v_pump_photon} the pump photon wavelength and signal photon wavelength, and likewise, the pump photon wavelength and idler photon wavelength. We observe that the reconstruction of the pump photon wavelength is dependent on the wavelength of down-converted photons, as expected from theory.

\section{Conclusion}

We demonstrated how to make a spectrometer with a pixel-based time-stamping fast camera, Tpx3Cam. Using it, we analyzed the spectral properties of single photons generated by a commercially-available Spontaneous Parametric Down-Conversion (SPDC) source. The calibration procedure is discussed, and can also be performed with other atoms instead of argon. Additionally, we note that our spectrometer works for a broadband range of wavelengths, covering the visible and near infrared parts of the electromagnetic spectrum. The key components that in principle can limit the range of wavelengths for our spectrometer are the diffraction grating, the camera, and the intensifier, however, all these components are broadband as discussed before. We can also can select parts of the spectrum by moving the camera position in a translation stage.

Knowing the spectral properties of the single photons from our source using this setup was proven to be used useful when we tested another spectrometer \cite{fast_spectrometer}. Note that in the present work, we employed a 2D pixel array, which makes the alignment of the setup easier. Examples of spectrometers using a 1D pixel array can be found in references~\cite{fast_spectrometer, Jennewein_spectr2014SR,Jennewein_spectr2014JAP, Lubin2021}.

The results presented here are valuable to other research groups interested in using the same commercial SPDC source, or in replicating this simple but useful spectrometer design for their needs.

\begin{acknowledgments}
This work was supported by the U.S. Department of Energy QuantISED award and BNL LDRD grant 22-22. B.F. and G. M. acknowledge support under the Science Undergraduate Laboratory Internships (SULI) Program by the U.S. Department of Energy. We are grateful to Michael DeArmond from Thorlabs and Thomas Tsang for helpful discussions on the SPDC source, and to Duncan England, Yingwen Zhang, and Michael Keach for help with the spectrometer design and implementation.
\end{acknowledgments}

\section*{Data Availability Statement}
The data that support the findings of this study are available from the corresponding author upon reasonable request.
\\
\\


\appendix

\bibliography{aipsamp}

\end{document}